# Experimental Electronic Structure of the Metallic Pyrochlore Iridate $Bi_2Ir_2O_7$

Q. Wang,[1,2] Y. Cao,[1] X.G. Wan,[3] J. D. Denlinger,[4] T. F. Qi,[5] O. B. Korneta,[5] G. Cao,[5] and D. S. Dessau[1]

[1]Department of Physics, University of Colorado, Boulder, CO 80309-0390, USA

[2]Los Alamos National Laboratory, Los Alamos, NM 87545, USA

[3]National Laboratory of Solid State Microstructures and Department of Physics, Nanjing University, Nanjing 210093, China

[4]Advanced Light Source, Lawrence Berkeley National Laboratory, Berkeley, California 94720, USA

[5]Center for Advanced Materials, Department of Physics and Astronomy, University of Kentucky, Lexington, KY 40506, USA

Angle-resolved photoemission measurements have been performed on $Bi_2Ir_2O_7$ single crystals, a prototypical example of the pyrochlore iridates. The density of states, the Fermi surface, and the near Fermi level band dispersion in the plane perpendicular to the (111) direction were all measured and found to be in overall agreement with our LDA + SOC density functional calculations. Our observations indicate the general validity of the LDA + SOC-based approach for the electronic structure of pyrochlore iridates, raising the possibility that some of the novel predicted phases such as quantum spin ice or Weyl Fermion states may exist in this family of compounds.

Recently, the 5d transition-metal-oxides (TMOs) have attracted significant attention due to their unique properties: they exhibit both strong electron correlation expected of d-electron systems and strong spin-orbit coupling (SOC) typical of high Z elements. The delicate interplay between electron correlation, SOC, inter-site hopping and crystal field splitting can lead to strongly competing ground states for these materials. The iridate compounds are important members of the 5d TMOs, for which a great amount of exotic phenomena have been observed experimentally or proposed theoretically, including spin-liquid state[1], $J_{eff}$ = 1/2 Mott insulator[2,3], correlated topological insulator[4], and high-$T_c$ superconductivity[5], and this list still keeps growing. Among iridate materials, the pyrochlore iridates $R_2Ir_2O_7$ (R = lanthanide element, Y, or Bi) are especially interesting due to the coexistence of the 5d-electron Ir subsystem[6,7], the 4f-electron Lanthanide magnetic moments[8,9], and the possible f-d exchange interaction between them[10]. The complicated physics in pyrochlore iridates leads to predictions for various novel phases such as quantum spin ice[8], topological insulator[11,12,13], correlated topological insulator[6], and Weyl Fermion state[7].

As a prototypical example of the pyrochlore iridates, $Bi_2Ir_2O_7$ (Bi227) is unique due to the lack of a magnetic moment in $Bi^{3+}$ ion, which offers an opportunity to distinguish the exotic physics in pyrochlore iridates originated from the Ir subsystem from the properties due to the other origins. The recent transport measurement on Bi227[14] revealed certain unconventional properties at low temperature and the muon spin relaxation study of Bi227[15] suggested two weak magnetic transitions at 1.84 and 0.23 K. All these results point to an exotic ground state in Bi227 due to the delicate interplay between Coulomb interaction and SOC of 5d electrons. Although there is great interest and extensive research effort on Bi227 and other pyrochlore iridates, their detailed experimental electronic structures are still lacking.

In this letter, we report the first ARPES study on a pyrochlore iridate. The near $E_F$ DOS was obtained and the result is fully in line with the $J_{eff}$ = 1/2 and 3/2 multiplet picture predicted by the local density approximation (LDA) + SOC calculations. The Fermi surface (FS) and near Fermi level ($E_F$) band dispersion in the plane perpendicular to (111) direction in momentum space were also obtained and are in overall agreement with our LDA + SOC calculations. Those observations indicate the general validity of the LDA + SOC-based approach for the electronic structure of pyrochlore iridates and also puts strong constraints on theories for describing the pyrochlore iridates. With the general agreement of the measured and calculated electronic structure of these materials, it becomes much more plausible that similar calculations of other closely related pyrochlore iridates will also be accurate. This increased the probability that some of the exotic physics predicted in these compounds will be accurate, for example the Weyl points and Fermi arcs in $Y_2Ir_2O_7$[7].

Bi227 single crystals were grown from off-stoichiometric quantities of $IrO_2$ and $Bi_2O_3$ using self-flux techniques in an oxygen rich atmosphere[14]. The ARPES experiments were performed at Beamline 4.0.3 (MERLIN) at the Advanced Light Source (ALS), Berkeley, in an ultra-high vacuum better than $3 \times 10^{-11}$ torr. The crystals were cleaved in situ, exposing a fresh and clean surface for the ARPES measurements. However, due to the lack of a natural cleavage plane the exposed surfaces were not beautiful mirrors, though clearly dispersive features were nevertheless observed. The angular resolution of the experiments was approximately $0.1^{o}$ and the energy resolution was 20~35 meV (depending upon photon energy).

Figure 1(a) shows the angle-integrated valence band spectrum of metallic Bi227 and insulating $Na_2IrO_3$ (Na213). Consistent with the metallic behavior reported by the transport measurement[14], the measured DOS of Bi227 exhibits a clear Fermi cutoff with multiple peaks at deeper binding energy (BE). As a general feature of the electronic structure of iridates, the near $E_F$ states of iridates are dominated by the Ir 5d-$t_{2g}$ orbitals and the O 2p orbitals. For example, the measured DOS of insulating Na213 contains two broad spectral features that are attributed to be the Ir 5d-$t_{2g}$ orbitals (-3 eV to $E_F$, with ~ 0.6 eV energy gap) and the O 2p orbitals (below -3 eV) by comparing with the theoretical calculations[16]. The DOS obtained from Bi227 shows a very similar distribution to that from Na213, except that the DOS of metallic Bi227 contains strong spectral weight at $E_F$. Thus the two broad spectral features observed in Bi227 could be naturally assigned to be the Ir $t_{2g}$ and O 2p orbitals, as indicated in figure 1(a).

In our measured Ir $t_{2g}$ DOS, there are clearly multiple spectral features which could be fitted into three Gaussian peaks plus a background, as shown in figure 1(b)[17]. The background may be due to a combination of the tail of the strong O 2p band extending towards $E_F$, an incoherent portion of spectral weight due to strong interactions, and the inelastic loss of energy that a small fraction of photoelectrons experience as they travel to the surface[18]. Similar to other correlated electron systems such as the cuprates and manganites, this background is relatively strong, but also similar to them it is difficult to determine the exact nature of its origin, so here we focus on the coherent peaks. The peak #1 from our fitting result is centered very near the Fermi energy and extends to about 0.5 eV below $E_F$, producing the metallic state for this compound. The peaks #2 and #3 are at deeper BE and are centered at approximately 0.7 eV and 1.4 eV, respectively. All of these results are in-line with our LDA + SOC calculation on Bi227[14], which also shows 3 peaks in this energy range (panel c, color-coded to match the data of panel b). In particular, the calculation shows three resolvable sets of spectral weight with about the same energy scale as observed in the data – one set (blue) centered near $E_F$ from the $J_{eff}$ = 1/2 bands, one $J_{eff}$ = 3/2 set centered near 0.7 eV (green), and one $J_{eff}$ = 3/2 set centered near 1.4 eV (yellow). Experimentally it is unclear whether a fourth set of spectral weight near 2.1 eV doesn't exist or if this is just buried under the tail of the large spectral weight originating from the O 2p band at higher energy. The similar separation of the $J_{eff}$ = 1/2 and $J_{eff}$ = 3/2 states indicates that the SOC parameter in the real material is very similar to that in the calculation, consistent with the concept that the SOC drives much of the physics of these materials. The peaky DOS at $E_F$ of $J_{eff}$ = 1/2 orbitals is also fully in line with the recent transport measurements on Bi227, in which the strong temperature-dependent behavior was qualitatively explained by the densely populated near $E_F$ electrons modulated by the thermal smearing effect[14].

Figures 2(a)-2(h) show the measured k-dependent intensity maps of Bi227 at different BE from $E_F$ to 1.1 eV. As previously noted, there is no obvious cleavage plane in Bi227 crystals. Thus in our experiment, the orientation of the exposed surface following the cleavage was determined spectroscopically. Since the intensity maps presented in figures 2(a)-2(h) exhibit a clear 6-fold symmetry pattern, this strongly indicates that the cleavage plane in our experiment is normal to the (1,1,1) direction of crystal. This cleavage plane orientation is further confirmed by the comparison between the experimental dispersion and theoretical calculation (discussed later). In figure 2(i), the Brillouin zone (BZ) for Bi227 with space group Fd-3m is shown. The blue hexagon represents the 2D Brillouin zone of the projected (1,1,1) surface with high-symmetry points $\bar{\Gamma}$, $\bar{K}$, and $\bar{M}$ labeled. This is the 2D Brillouin zone that we use here, shown as the grey dashed lines in figures 2(a)-2(h). From figure 2(a), a pocket-like FS is clearly resolved around the BZ center. Based on the intensity maps from $E_F$ to 0.3 eV, it is hard to tell if this pocket is electron-like or hole-like. Later we will show that it is actually a hole-like pocket. The intensity map at BE = 0.7 eV shows strong spectral weight at the zone center, with that spectral weight slowly spreading out at deeper BE. This indicates another hole-like dispersion with band top at BE ~ 0.7 eV.

A more detailed analysis on the Bi227 FS and dispersions along the high symmetry directions is presented in figure 3. Figure 3(a) shows the FS of Bi227 again but taken with smaller angular steps and better statistics. To acquire accurate positions of Fermi crossings, the momentum-distribution-curves (MDC) at $E_F$ have been fitted using Lorentzian lineshapes plus a linear background, and the fitted peak locations that correspond to the Fermi crossing positions are shown as the green marks in figure 3(a). The fitting result shows a clear rounded-triangle-shape FS, consistent with the symmetry properties of the 2D Brillouin zone of the projected (1,1,1) surface (red boxes in panel (a)). Figures 3(b1)-3(b6) show the spectra taken along cuts C1-C6 as indicated by the yellow cuts in panel (a). Figures 3(c1)-3(c6) are the second-derivative images along the momentum direction of spectra (b1)-(b6), respectively, which enhances the contrast of the raw spectra and makes is easier to track the electronic dispersion.

Consistent with the DOS presented in figure 1, the angle-resolved spectra shown in panels (b1)-(b5) contain two sets of spectral features: the dispersive feature from $E_F$ to BE ~ 0.5 eV with relatively weak spectral weight which corresponds to the $J_{eff}$ =1/2 doublet bands (blue peak #1 in figure 1(b)), and the dispersive feature below 0.5 eV with relatively strong spectral weight which corresponds to the upper portion of the $J_{eff}$ =3/2 quartet bands (peak #2 figure 1(b)). Due to the weak spectral weight of the dispersive feature from $E_F$ to 0.5 eV, the band dispersion associated with the Fermi pocket shown in figure 3(a) cannot be fully resolved from the raw spectra. But from the 2$^{nd}$-derivative images as shown in panels (c1)-(c6), there are bands clearly dispersing away from the zone center, which indicates hole-like dispersion associated with the Fermi pocket. As for the dispersive feature below 0.5 eV, both raw spectra and second-derivative images indicate a very broad hole-like dispersion from BE ~ 0.7 eV, which is also consistent with the intensity maps from 0.7 eV to 1.1 eV presented in figure 2.

To further investigate the electronic structure of Bi227, we have performed band calculations based on the local density approximation to density functional theory (DFT) using the full-potential, all-electron, linear-muffin-tin-orbital (LMTO) method[19]. The 3-dimensional FS and the corresponding dispersions are obtained from the non-magnetic LDA + SOC calculation. To make a direct connection to our experimental observations, the calculated FS and dispersions in the plane perpendicular to (1,1,1) direction and centered at different locations along (1,1,1) line are compared to our ARPES result. The best overall agreement is obtained between the observed electronic structure and the calculated electronic structure in the plane centered at (1/4,1/4,1/4) point [20] (the middle point between Γ and L shown in figure 2(i)) and perpendicular to (1,1,1) direction in momentum space. Those calculated FS and dispersions along high symmetry directions are plotted on top of the observed spectra, as shown in figures 3(a), 3(c1), and 3(c6).

In figure 3(a), both the observed (green marks) and calculated (black dotted line) FS show a round-triangle-like pocket with similar size. Consistent with the ARPES result, the calculated dispersion (yellow lines) along the high symmetry directions shown in figures 3(c1) and 3(c6) also indicate hole-like dispersion associated with the Fermi pocket with similar position of Fermi crossing. Below $E_F$, the calculated dispersion contains a flat band at BE ~ 0.3 eV. This band is not well resolved in figures (c1) and (c6) since the $2^{nd}$-derivative images along momentum direction cannot emphasize the weak dispersive features in energy direction. On the other hand, both raw spectra shown in figures 3(b1) and 3(b6) and intensity map shown in figures 2(d) indicate certain spectral weight at BE ~ 0.3 eV near zone center, which is consistent with this flat band in calculation. As for the dispersive feature at deeper binding energy, corresponding to the strong spectral weight observed below 0.5 eV, the calculation indicates multiple bands in the same energy range. Though the detailed dispersion for each band is not strictly hole-like, the overall trend of the dispersion below ~ 0.8 eV indeed show hole-like feature, which is consistent with the overall spectral weight distribution observed in the experiment.

Here we note that the calculation discussed above does not include correlation U. It has been shown that the strong hybridization between Ir 5d and Bi 6p electrons and the screening effect will reduce the correlation effect in Bi227[14] and always lead it to a metallic state. The overall agreement between our ARPES result and the non-magnetic calculation is fully in line with the small effective correlation in Bi227. Furthermore, the excellent agreement between experimental observation and the LDA calculation in Bi227 indicates the general validity of the LDA + SOC based approach for the electronic structure of pyrochlore iridates. This will also strengthen the case that other LDA + SOC-based expectations on related compound may be realistic, e.g., the topological semimetal in $R_2Ir_2O_7$ (R = lanthanide element or Y) type pyrochlore iridates, predicted by the LDA + SO + U calculations[10].

To summarize, the electronic structure of Bi227 was obtained by using ARPES. The DOS measured near $E_F$ is fully in line with the $J_{eff} = 1/2$ and $3/2$ multiplets picture predicted by the LDA + SOC calculations. The Fermi surface (FS) and near Fermi level ($E_F$) band dispersion in the plane perpendicular to (111) direction in momentum space were obtained and are in overall agreement with the available LDA + SOC calculations, even without the addition of an on-site correlation term U. Our observations indicate the general validity of the LDA + SOC-based approach for the electronic structure of pyrochlore iridates, indicating that such success is also likely to carry through to other members of the pyrochlore iridates (e.g. obtained by replacing Bi with other elements such as Y or Gd) in which novel behavior such as unusual topological states, Weyl Fermion states, or spin-ice, are predicted and also puts strong constraints on theories for describing the pyrochlore iridates.

This work was supported by the National Science Foundation under grant DMR-1007014 to the University of Colorado and grants NSF DMR-0856234, NSF DMR-1265162, and EPS-0814194 to the University of Kentucky. This work is also based in part upon research conducted at the Advanced Light Source, which is funded by the US Department of Energy.

**Fig. 1.**

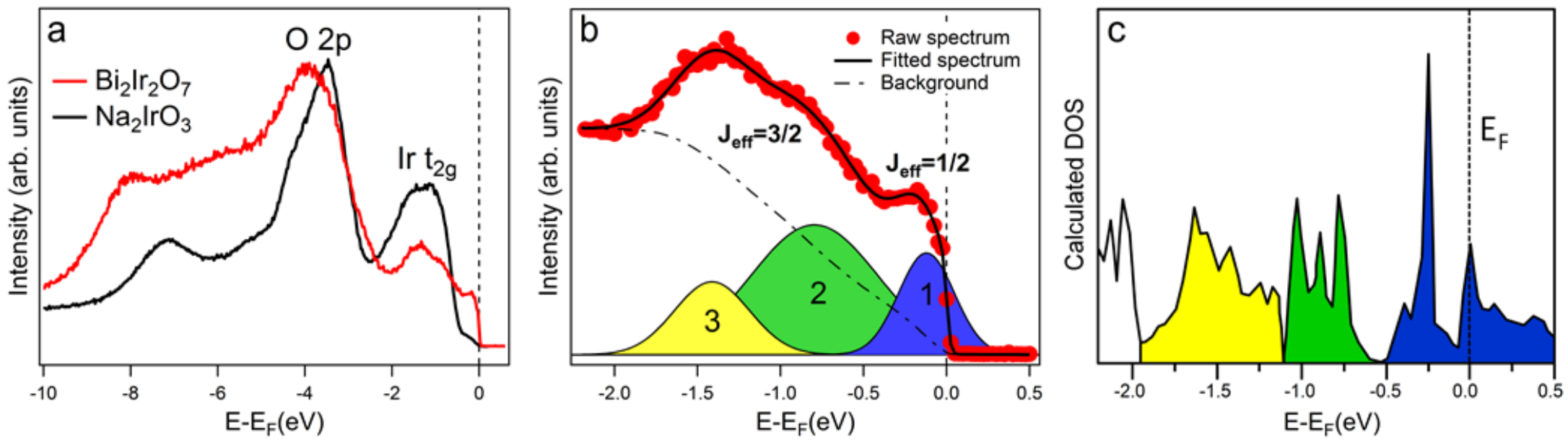

FIG. 1: (a) Angle-integrated photoemission spectrum of $Bi_2Ir_2O_7$ and $Na_2IrO_3$. (b) Model fit of the Ir $t_{2g}$ valence band with 3 Gaussian peaks, a background and Fermi function. (c) Calculated density of states of $Bi_2Ir_2O_7$ from ref. 14, color-coded to match the peaks of panel b. The $Bi_2Ir_2O_7$ data was taken with 100 eV photons at 25 K. The $Na_2IrO_3$ data was taken with 100 eV photons at 150K.

**Fig. 2.**

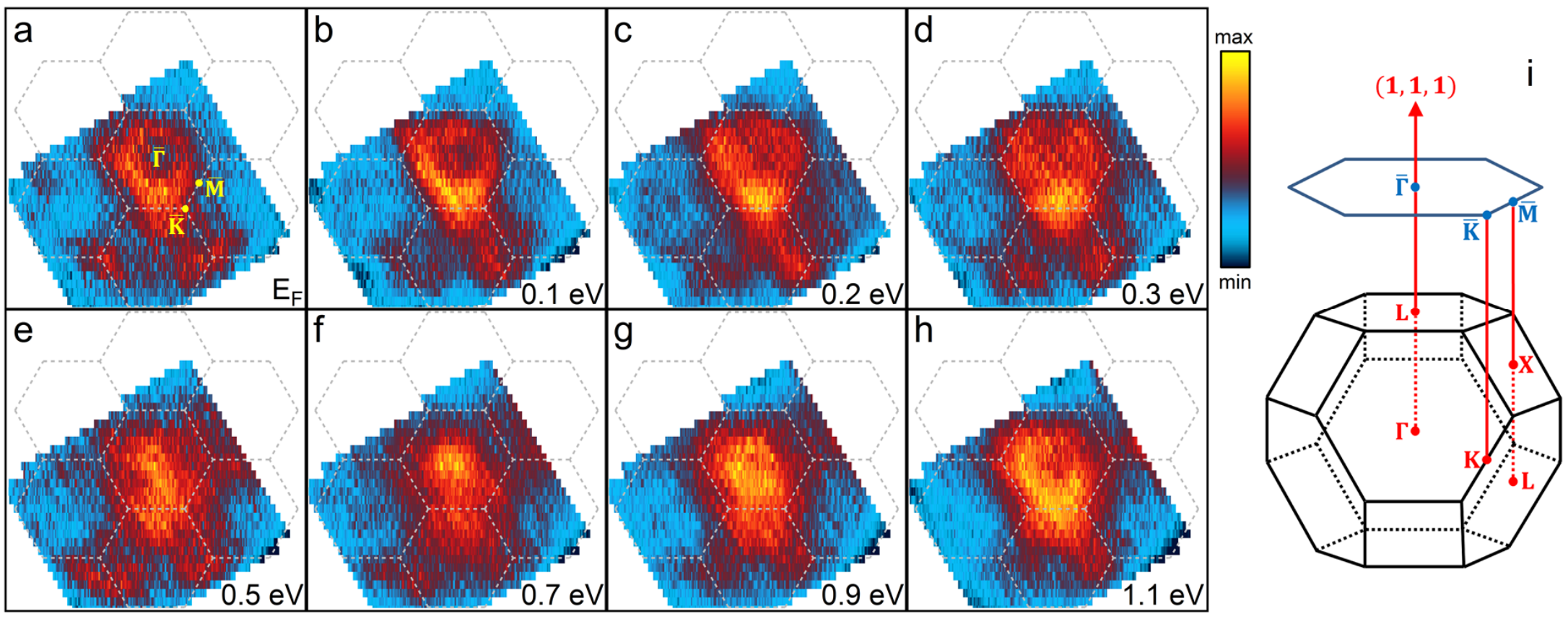

Fig. 2: (a)-(h) Intensity maps of $Bi_2Ir_2O_7$ at different binding energies from $E_F$ to 1.1 eV, which show a clear 6-fold symmetry pattern and indicate the cleavage plane is normal to the (1,1,1) direction of the crystal. The grey dashed lines represent the projected 2D surface BZ normal to the (1,1,1) direction, as indicated in panel (i). The data was taken with 80 eV photons at 25 K.

# Fig. 3.

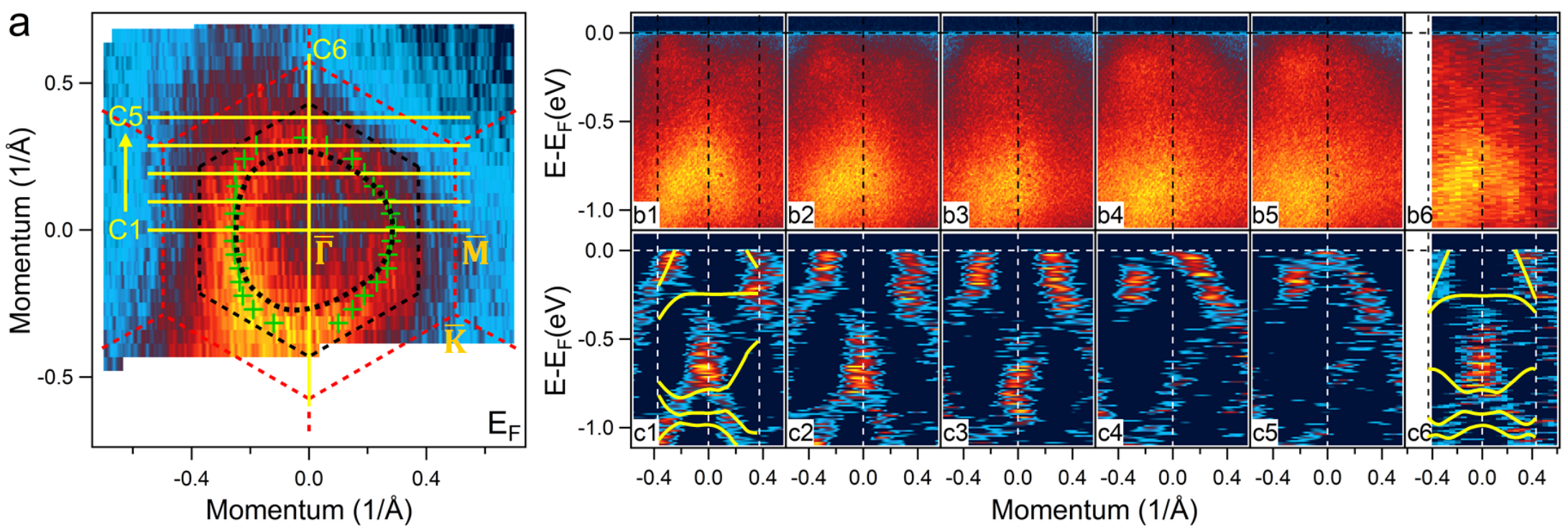

Fig. 3. (a) Detailed Fermi surface of $Bi_2Ir_2O_7$. The green marks show the positions of Fermi crossing which are obtained by fitting the MDC peaks at the Fermi level. The black dotted circle represents the calculated FS in the plane centered at (1/4,1/4,1/4) point and perpendicular to (1,1,1) direction in momentum space. The red dashed boxes indicate the projected surface BZ boundary, while the black dashed box indicates the bulk BZ boundary perpendicular to (1,1,1) direction and centered at L point, as shown in fig. 2(i). Panels (b1)-(b6) are spectra taken along cut C1-C6 as shown in panel (a). Panels (c1)-(c6) are the corresponding $2^{nd}$-derivative images along momentum direction (MDC). The calculated dispersions along high symmetry directions are shown as the yellow lines in panels (c1) and (c6). The data was taken with 80 eV photons at 25 K.